\documentclass[11pt,showpacs,preprintnumbers,superscriptaddress,amsmath,amssymb,nofootinbib]{revtex4-1}
\usepackage{graphicx}
\usepackage{dcolumn}
\usepackage{bm}
\usepackage{amssymb}
\usepackage{amsmath}
\usepackage{epsfig}    
\usepackage{color}
\usepackage{slashed}
\usepackage{hhline}
\usepackage[normalem]{ulem}
\usepackage{mathrsfs} 
\def\be{\begin{equation}}
\def\ee{\end{equation}}
\def\Mat3#1#2#3#4#5#6#7#8#9{
  \left(
  \begin{array}{ccc}
    #1 & #2 & #3 \\
    #4 & #5 & #6 \\
    #7 & #8 & #9 \\
  \end{array}
  \right) }
\newcommand{\bea}{\begin{eqnarray}}
\newcommand{\eea}{\end{eqnarray}}
\newcommand{\nn}{\nonumber}

\numberwithin{equation}{section}

\usepackage{slashed,color,graphicx}
\usepackage{hyperref}
\hypersetup{
  colorlinks = true,
  linkcolor = blue,
  citecolor = magenta
} 

\begin{document}

%%%%%%%%%
% \title{Lepton? anomalous magnetic moment and inverse seesaw in gauged $U(1)_{B-L}$ extended two-Higgs doublet model with type-X}
\title{Inverse seesaw and $(g-2)$ anomalies in $B-L$ extended two Higgs doublet model}
\preprint{KIAS-P21010, APCTP Pre2021-004}

\author{Tanmoy Mondal}
\email{tanmoy@kias.re.kr}
\affiliation{School of Physics, KIAS, Seoul 02455, Korea}
\author{Hiroshi Okada}
\email{hiroshi.okada@apctp.org}
\affiliation{Asia Pacific Center for Theoretical Physics (APCTP) - Headquarters San 31, Hyoja-dong,
  Nam-gu, Pohang 790-784, Korea}
\affiliation{Department of Physics, Pohang University of Science and Technology, Pohang 37673, Republic of Korea}

\date{\today}

\begin{abstract}
We propose a gauged $U(1)_{B-L}$ extended two Higgs doublet model to explain both neutrino mass and lepton anomalous magnetic 
moments ($g-2$). Neutrino mass is generated via an inverse seesaw mechanism by introducing singlet fermions.
%%%
Especially, we update the result of muon $g-2$ in light of the very recent report by  E989 experiment at Fermilab, indicating
$a_\mu^{\rm FNAL}=116592040(54)\times 10^{-11}$.  Combining BNL result, we have the following deviation from the standard model prediction $\Delta a_\mu=(2.51\pm 5.9)\times10^{-10}$ at 4.2 $\sigma$.
%%%
Thanks to an appropriate assignment for $U(1)_{B-L}\times Z_2$ symmetry and larger $(20\lesssim)\tan\beta$ that is favoured by 
type-X model, we realize natural hierarchies among neutral fermions. The lepton anomalous magnetic moments can be induced at 
the one loop level by introducing an iso-spin singlet singly-charged boson.  This charged scalar plays a significant role in 
evading chiral suppression of these phenomenologies. We show sizable lepton ($g-2$) can be obtained after satisfying all the 
flavour constraints, such as $\mu\to e\gamma$ and flavour conserving leptonic $Z$ boson decays.
\end{abstract}

\maketitle
\newpage

\section{Introduction}
The two Higgs doublet models (THDMs) are one of the minimal scenarios beyond the Standard Model (SM) to extend the Higgs sector at the electroweak (EW) scale. Numerous intriguing phenomenologies are inherent in these models, which are absent in the iso-spin singlet extended models. Several new particles might be found by detecting non-standard signatures at the Large Hadron Collider (LHC) in the near future, which would naturally be the next step after the discovery of the SM Higgs boson with 125 GeV mass.

Even though there typically exist four types of THDMs~\cite{Branco:2011iw}, the type-X or lepton specific scenario is one of the successful scenarios to explain the muon anomalous magnetic moment ($g-2$).
It is possible to satisfy all the constraints such as EW precision test especially about the pseudo-scalar Higgs, vacuum stability, perturbativity with direct and indirect bounds from collider and B physics~\cite{Cao:2009as,Broggio:2014mna,Wang:2014sda,Ilisie:2015tra,Abe:2015oca,Han:2015yys,Chun:2016hzs,Cherchiglia:2016eui,Cherchiglia:2017uwv,Wang:2018hnw, Li:2018aov}. 
 
It is possible to explain electron ($g-2$) in addition to muon ($g-2$) in modified type-X by introducing vector like leptons~\cite{Chun:2020uzw,Hernandez:2021tii}\footnote{See also, \cite{Barman:2018jhz,Frank:2020smf,Dermisek:2021ajd, Jana:2020pxx, Jana:2020joi, Dermisek:2020cod, Crivellin:2018qmi, Crivellin:2015hha}}. 
Then, sizable the electron ($g-2$) is achieved by mediating the exotic fermions at a one-loop level, while the muon ($g-2$) can be explained at the two-loop level.

Inverse seesaw is a low scale seesaw mechanism that explains the neutrino mass generation by introducing a pair of singlet fermions~\cite{Mohapatra:1986bd}.
The Dirac neutrino mass term in the inverse seesaw scenario can be quite large since the smallness of neutrino mass originated from lepton number violating $\mu$ 
term which is \emph{naturally} small. The chirality flipping Dirac mass term can enhance the lepton ($g-2$) contribution in these scenarios.

In this paper, we introduce neutral fermions in type-X THDM scenario to explain neutrino mass via inverse seesaw~\cite{Wyler:1982dd,Mohapatra:1986bd,Khalil:2010iu}
mechanism. The gauged $U(1)_{B-L}$ symmetry is responsible for small lepton number violating mass term, and natural hierarchies among neutral fermion mass matrices are realized by higher dimensional terms and large $\tan\beta$. The charged scalars in the THDM and the fermionic fields necessary for the inverse seesaw mechanism can introduce a new contribution to lepton ($g-2$), albeit small due to chiral suppression. We introduced a singlet charged scalar to mitigate the issue.  Singly-charged boson plays a crucial role in evading the chiral suppression of the lepton ($g-2$)~~\cite{Lindner:2016bgg}~\footnote{There exists another way to evade the chiral suppression by introducing singly-charged  fermions~\cite{Okada:2013iba, Baek:2014awa, Okada:2014nsa}.}.
We show that the model can explain the lepton ($g-2$) anomalies without violating constraints coming from the lepton flavour violation decays, 
by demonstrating a benchmark point with proper inputs and outputs in both the cases of normal and inverted hierarchies of neutrino mass ordering. The values of lepton $(g-2)$ are within the range of 2$\sigma$ confidence level (C.L.), and the masses for neutral heavy fermions and singly-charged bosons are heavy enough to evade the collider bounds.
 %%%
% Notice here that larger $\tan\beta$, {which should be more 5 at minimum value}, is allowed in type-X model to satisfy B-physics constraints such as $b\to s\gamma$ and $b\to s\ell\bar\ell$.

This paper is organized as follows.
In Sec.~II, we review our model, showing our relevant Lagrangian, Higgs potential, and neutral fermions including the active neutrino masses, and formulating the lepton ($g-2$), flavor conserving leptonic $Z$ boson decays. In Sec.III we demonstrate our numerical analysis and we will show a promising benchmark point at 1$\sigma$ interval of muon ($g-2$) which satisfy neutrino as well as lepton flavor violation data. In Sec.~IV, we devote the conclusions and discussions. 
%In Appendix, we review...
%%%%%%%%%%%%%%%%%%%%%%%%%%%%%%%%%%%%%
% \subsection{Particle contents and the model Lagrangian}
\begin{widetext}
  \begin{center} 
    \begin{table}%[tbc]
      \begin{footnotesize}
        \begin{tabular}{|c||c|c|c|c|c|c|c||c|c|c|c|c|c|}\hline 
          & \multicolumn{7}{c||}{Fermions} & \multicolumn{4}{c|}{Bosons}\\\hline\hline 
          Symmetry  ~&~ $ Q_{L}$ ~&~ $u_{R}$ ~&~ $d_R$ ~&~ $L_L$   ~&~ $e_R$  ~&~ $N_L(N'_L)$ ~&~ $ N_R$ ~ &~ $H_1$  ~   &~ $H_2$ ~ &~ $\varphi$~  &~ $\chi^-$~
          \\\hline 
          $SU(3)_C$  & $\bm{3}$  & $\bm{3}$  & $\bm{3}$ & $\bm{1}$  & $\bm{1}$ & $\bm{1}$     & $\bm{1}$  & $\bm{1}$  & $\bm{1}$    & $\bm{1}$ & $\bm{1}$  
          \\\hline 
          %%%
          $SU(2)_L$  & $\bm{2}$  & $\bm{1}$  & $\bm{1}$   & $\bm{2}$  & $\bm{1}$   & $\bm{1}$  & $\bm{1}$ & $\bm{2}$  & $\bm{2}$    & $\bm{1}$  & $\bm{1}$    
          \\\hline 
          %%%
          $U(1)_Y$  & $\frac{1}{6}$ & $\frac23$ & $-\frac13$  & $-\frac23$ &  $-1$  &  $0$  & $0$ & $\frac12$    & $\frac12$  & $0$      & $-1$
          \\\hline
          %%%
          $U(1)_{B-L}$ & $\frac13$  & $\frac13$ & $\frac13$ & $-1$  & $-1$   & $-\frac12(\frac12)$ & $-1$   & $0$   & $0$ & $\frac12$   & $-\frac12$
          \\\hline
          %%%
          $\mathbb{Z}_2$  & $+$  & $+$ & $+$ & $+$ & $-$ & $+(-)$ & $+$ & $-$     & $+$   & $+$ & $-$     \\\hline
          %%%
          %$\mathbb{Z}_2$ & $+$   & $-$  & $+$ & $+$& $-$& $-$& $+$ & $+$  \\\hline\hline
        \end{tabular}
        \caption{Field contents of the fermions
          and their charge assignments under $SU(3)_C\times SU(2)_L\times U(1)_Y\times  U(1)_{B-L} \times Z_2$,
          where family indices are abbreviated.}
        \label{tab:1}
      \end{footnotesize}
    \end{table}
  \end{center}
\end{widetext}

% 
% \begin{table}[t]
% \centering {\fontsize{10}{12}
% \begin{tabular}{|c||c|c|c|c|c| }\hline\hline
% %&\multicolumn{1}{c||}{VEV$\neq 0$} & \multicolumn{3}{c|}{Inert } \\\hline
%   Bosons  &~ $H_1$  ~   &~ $H_2$ ~ &~ $\varphi$~  &~ $\chi^-$~   \\\hline
% $SU(3)_C$ & $\bm{1}$  & $\bm{1}$    & $\bm{1}$ & $\bm{1}$        \\\hline 
% $SU(2)_L$ & $\bm{2}$  & $\bm{2}$    & $\bm{1}$  & $\bm{1}$        \\\hline 
% $U(1)_Y$ & $\frac12$    & $\frac12$  & $0$      & $-1$     \\\hline
%  $U(1)_{B-L}$  & $0$   & $0$ & $\frac12$   & $-\frac12$    \\\hline
% $Z_2$ & $+$     & $-$   & $+$ & $-$  \\\hline
% \end{tabular}%
% } 
% \caption{Field contents of bosons
% and their charge assignments under $SU(3)_C\times SU(2)_L\times U(1)_Y\times  U(1)_{B-L} \times Z_2$. }
% \label{tab:2}
% \end{table}
% 
% %\subsection{Lagrangian}

 \section{Model setup}
In this section, we review our model set up under the SM and $U(1)_{B-L}\times Z_2$ symmetries.  In addition to the SM model fields, we introduce three neutral heavy fermions $N\in (N_L,N'_L,N_R)$ in the fermion sector.  Three families of $N_R$ are requested to have anomaly cancellations in the gauged $U(1)_{B-L}$ symmetry.  While only the two families of $N_L, N'_L$ are needed to explain the neutrino oscillation data, but we introduce three families to match the $N_R$s. Note here that the number of $N'_L$, which has opposite charges of $N_L$ under $U(1)_{B-L}\times Z_2$,  should be the same one of $N_L$ due to the anomaly cancellation. 
 In addition to the $U(1)_{B-L}$, we impose $\mathbb{Z_2}$ symmetry, and we assign $e_R$ to be odd under this symmetry while other SM fermions are even. 
 All the charge assignments for fermions are summarized in Table~\ref{tab:1}.
 Apart from the SM fermions, we also assumed that the $N_L'$ is  $\mathbb{Z}_2$ odd to remove $N_L'\varphi N_R$  interaction term, which can spoil the inverse seesaw structure as we will discuss later. 

In the bosonic sector, we introduce an isospin doublet boson $H_1$
and  two isospin singlets $\varphi$ and $\chi^-$ in addition to the SM boson $H_2$, where only $H_1$ and $\chi^-$ have minus sign under $\mathbb{Z}_2$ .  The charge assignment of $H_2$ and $e_R$ ensures that no tree-level flavour changing neutral current interactions are possible, and the respective Yukawa structure is called type-X THDM. The singlet boson $\varphi$ introduce the Dirac mass term of $N$ and Majorana terms of $N_L^{(')}$ after breaking the $U(1)_{B-L}$ symmetry spontaneously. It also contributes to the mixing between the physical singly charged bosons of $\chi^-$ and the charged-component of the doublet scalars. Notice that the Dirac mass terms appear at tree level, while Majorana masses of $N_L^{(')}$ are five-dimensional, which might explain the smallness of the Majorana mass term. The singly-charged boson $\chi^-$ plays a crucial role to explain the muon anomalous magnetic moment at the one-loop level as discussed later. All the charge assignments for bosons are summarized in Table~\ref{tab:1}. 

Under these symmetries, one writes the valid Lagrangian as follows:
\begin{align}
  -\mathcal{L}_Y & = 
  y_u \bar Q_{L} \tilde H_2^* u_R +y_d   \bar Q_{L}  H_2 d_R + f \bar L_{L}  \tilde H_2^*  N_R + y_\ell \bar L_{L}  H_1 e_R 
  + g \bar N_L e_R \chi^+
  \nn\\
  &+y_{N}\bar N_L N_R\varphi + \frac{\lambda_{L}}{\Lambda}\bar N_L^C N_L\varphi^2 + \frac{\lambda_{L'}}{\Lambda}\bar N^{'C}_L N'_L\varphi^{*2}
+  {\rm h.c.} ,
\label{yukawa}
\end{align}
where $\tilde H_{1,2}^*\equiv i\sigma_2 H_{1,2}^*$, $\sigma_2$ is the second component of the Pauli matrix, and the charged-lepton sector is assumed to be diagonal at the flavor level for simplicity.
$N'$ has mass at 5 dimensional term after spontaneously symmetry breaking under $U(1)_{B-L}$, which might be rather small~\cite{Bandyopadhyay:2012px}.
Even though it would be a good dark matter candidate~\cite{Abdallah:2019svm} with light mass scale that would be an interesting topic, we will leave it to our future's work.
%In our paper, we do not discuss this field hereafter, since it would not have any interesting interactions.
%%%
%%%
The Higgs potential is given by 
% \begin{align}
%   %%%
% &  {\cal V} \equiv     {\cal V}_{non-trivial}  +  {\cal V}_{trivial}  ,\\
%   %
% &   {\cal V}_{non-trivial}  = \lambda_0 (H^T_1 i\sigma_2 H_2)\chi^- \varphi +\lambda'_0 (H^\dag_1 H_2)^2
%  + {\rm h.c.}\\
% %
% &   {\cal V}_{trivial}  =\sum_{\Phi=H_1}^{H_2,\varphi,\chi^-} \mu_{\Phi} (\Phi^\dag\Phi)
% +\sum_{\Phi_1\le\Phi_2}^{H_1,H_2,\varphi,\chi^-} \lambda_{\Phi_1\Phi_2} (\Phi_1^\dag\Phi_2)^2
%  ,\label{Eq:pot}
% \end{align}
\begin{align}
  %%%
{\cal V} &\equiv     {\cal V}_{New}  +  {\cal V}_{2HDM}  , \label{Eq:pot}
\\
{\cal V}_{new}  &=\mu_\varphi^2 \varphi^* \varphi + \mu_\chi^2 \chi^+\chi^- + \lambda_\varphi \left(\varphi^*\varphi\right)^2 + \lambda_\chi \left(\chi^+\chi^-\right)^2 + \lambda_{\varphi\chi}\left(\varphi^*\varphi\right)\left(\chi^+\chi^-\right) \\\nonumber
& \sum_{\Phi=H_1,H_2}\left\{\lambda_{\Phi\varphi} \left(\varphi^*\varphi\right)\left(\Phi^{\dagger}\Phi\right) + \lambda_{\Phi\chi} \left(\chi^+\chi^-\right)\left(\Phi^{\dagger}\Phi\right) \right\} + \left\{ \lambda (H^T_1 i\sigma_2 H_2)\chi^- \varphi + h.c. \right\}\\
%
% \begin{eqnarray}
{\cal V_{\mathrm{2HDM}}} &= m_{11}^2\Phi_1^{\dagger}\Phi_1 + m_{22}^2\Phi_2^{\dagger}\Phi_2 -\Big[m_{12}^2\Phi_1^{\dagger}\Phi_2 + \mathrm{h.c.}\Big]
+\frac{1}{2}\lambda_1\left(\Phi_1^\dagger\Phi_1\right)^2+\frac{1}{2}\lambda_2\left(\Phi_2^\dagger\Phi_2\right)^2 \\
\nonumber & +\lambda_3\left(\Phi_1^\dagger\Phi_1\right)\left(\Phi_2^\dagger\Phi_2\right)+\lambda_4\left(\Phi_1^\dagger\Phi_2\right)\left(\Phi_2^\dagger\Phi_1\right)
+\frac{1}{2}\lambda_5\Big\{ \left(\Phi_1^\dagger\Phi_2\right)^2+\left(\Phi_2^\dagger\Phi_1\right)^2\Big\},
\end{align}

where we define $H_a\equiv [H^+_a,(v_a+ r_a + iz_a)/\sqrt2]^T$ and $\varphi \equiv  (v_{\varphi} +r' + iz')/\sqrt2$ (a=1,2).
Here, we assume that there is no mixing between $H_{1,2}$ and $\varphi$ which is favored by the current experiment at LHC.
Then, the part of $H_{1,2}$ is the same as the one of type-X~\cite{Branco:2011iw}, and we parameterize the following way: 
\begin{align}
  %%%
& H_1^0  = \frac{1}{\sqrt2}\left[ v_H c_\beta + c_\alpha H - s_\alpha h + i c_\beta G_0 - i s_\beta A \right]\\
&  H_2^0 = \frac{1}{\sqrt2 }\left[ v_H s_\beta + s_\alpha H + c_\alpha h +i s_\beta G_0+ i c_\beta A \right]
 ,\label{Eq:not-neut}
\end{align}
where $h$ is the SM Higgs boson and $H(A)$ is additional scalar (pseudo-scalar), {$v_1\equiv v_H c_\beta,\ v_2\equiv v_H s_\beta$,} and $v_H\equiv \sqrt{v^2_{1}+v^2_{2}}\approx 246$ GeV. The singly-charged bosons are diagonalized, $O_C M_C^2 O_C^T ={\rm diag}[0,m^2_{c_1},m^2_{c_2}]$, 
and given in terms of mass eigenstates as follows:
\begin{align}
  %%%
&
\left[\begin{array}{c}
H^-_1 \\ 
H^-_2 \\ 
\chi^-  \\ 
\end{array}\right]=
O_C^T
\left[\begin{array}{c}
c_0^-  \\ 
c_1^-  \\ 
c_2^- \\ 
\end{array}\right],\\
%\quad s_{2\gamma}\simeq \frac{\lambda_0 v_{H_2} v_{\varphi} }{m_{c_1}^2 - m_{c_2}^2},
& M^2_C =\frac12
\left[\begin{array}{ccc}
2 m_{12}^2\dfrac{v_2}{v_1}-(\lambda_4+\lambda_5) v_2^2~~ & -2 m_{12}^2+(\lambda_4+\lambda_5) v_1 v_2  &  \lambda v_2 v_\varphi  \\ 
* & 2 m_{12}^2\dfrac{v_1}{v_2}-(\lambda_4+\lambda_5) v_1^2 &  -\lambda v_1 v_\varphi \\ 
* & * & 2 \mu_\chi^2 +\left(\lambda_{H_1\chi} v_1^2 +  \lambda_{H_2\chi} v^2_2 + \lambda_{\varphi\chi} v_\varphi^2 \right)\\ 
\end{array}\right],
\label{Eq:not-sgly}
\end{align}
where $O_C O_C^T=O_C^T O_C=1_{3\times3}$ is the orthogonal mixing matrix, $c^-_{1,2}$ are mass eigenstates for singly-charged bosons, while $c_0^-$ is absorbed by singly-charged gauge boson $W^-$.
%Here, we define these mass eigenvalues to be $m^2_{c_{1,2}}$.
%~\footnote{In general, there exists mixing between $H_1^+$ and $H^+_2$, we simply assume to be no mixing of them, for the time being.}
%where we assume that  $H^+_2$ is the mass eigenstate after diagonalizing $H^+_1$ and $H^+_2$.

% \noindent \underline{\bf Neutrino sector}:\\
\subsection{Neutrino sector}
After the spontaneous symmetry breaking, the neutral fermion mass matrix based on $(\nu_L,N_R^C,N_L)^T$ is given by,
\begin{align}
M_N
&=
\left[\begin{array}{ccc}
0 & m_D^* & 0  \\ 
m_D^\dag & 0 & M^\dag \\ 
0  & M^* & \mu_L \\ 
\end{array}\right],
\end{align}
where $m_D\equiv f v_{H_1}/\sqrt2,\ M\equiv y_N v_{\varphi}/\sqrt2$ are diagonal while $\mu_L\equiv \frac{\lambda_L v_{\varphi}^2}{2\Lambda}$ is symmetric $3\times 3$ mass matrix.
%, and $\psi_R^{0c}$ and $\psi_L^0$ are neutral components of $\psi_R$ and $\psi_L$.
%%% Here we assume $\mu<< m_\ell,m_D$.
Then the active neutrino mass matrix can be given as
\begin{align}
m_\nu\approx m^*_D (M^*)^{-1} \mu_L (M^\dag)^{-1} m^\dag_D.
\end{align}
Once we fix $m_DM^{-1}\sim{\cal O}(10^{-3})$, then $\mu_L\sim {\cal O}(10^{-5})$ GeV in order to satisfy the observed neutrino mass squared differences.
The neutrino mass matrix is diagonalized by unitary matrix $U_{MNS}$; $D_\nu= U_{MNS}^T m_\nu U_{MNS}$, where $D_\nu\equiv {\rm diag}[m_1,m_2,m_3]$.
The $U_{MNS}$ matrix is parameterized by three mixing angle $\theta_{ij} (i,j=1,2,3; i < j)$, one CP violating Dirac phase $\delta_{CP}$, and two Majorana phases $\{\alpha_{21}, \alpha_{32}\}$ as follows:
\begin{equation}
U_{\mathrm{PMNS}} = 
\begin{pmatrix} c_{12} c_{13} & s_{12} c_{13} & s_{13} e^{-i \delta_{CP}} \\ 
-s_{12} c_{23} - c_{12} s_{23} s_{13} e^{i \delta_{CP}} & c_{12} c_{23} - s_{12} s_{23} s_{13} e^{i \delta_{CP}} & s_{23} c_{13} \\
s_{12} s_{23} - c_{12} c_{23} s_{13} e^{i \delta_{CP}} & -c_{12} s_{23} - s_{12} c_{23} s_{13} e^{i \delta_{CP}} & c_{23} c_{13} 
\end{pmatrix}
\begin{pmatrix} 1 & 0 & 0 \\ 0 & e^{i \frac{\alpha_{21}}{2}} & 0 \\ 0 & 0 & e^{i \frac{\alpha_{31}}{2}} \end{pmatrix},
\end{equation}
where $c_{ij}$ and $s_{ij}$ stands for $\cos \theta_{ij}$ and $\sin \theta_{ij}$ respectively. 

In our numerical analysis, we will adopt the best fit values of neutrino experimental data~\cite{Esteban:2018azc} for normal hierarchy (NH) and inverted hierarchy (IH) as
shown in Table~\ref{table:osc-param-max-min}. 
% \begin{align}
% &{\rm NH}: \Delta m^2_{\rm atm}=2.514\times 10^{-3}\ {\rm eV}^2,\
% \Delta m^2_{\rm sol}=7.42\times 10^{-5}\ {\rm eV}^2,\label{eq:nh}\\
% &\sin^2\theta_{13}=0.02221\ 
% \sin^2\theta_{23}=0.570,\ 
% \sin^2\theta_{12}=0.304,\ \delta_{CP}=195^{\circ}.\nn\\
% %%%
% &{\rm IH}: \Delta m^2_{\rm atm}=2.497\times 10^{-3}\ {\rm eV}^2,\
% \Delta m^2_{\rm sol}=7.42\times 10^{-5}\ {\rm eV}^2,\label{eq:ih}\\
% &\sin^2\theta_{13}=0.02240,\ 
% \sin^2\theta_{23}=0.575,\ 
% \sin^2\theta_{12}=0.304,\ \delta_{CP}=286^{\circ}, \nn
% \end{align}
% where we set to be zero for Majorana phases for simplicity.
%where NO and IO stand for normal and inverted ordering respectively.

%%%%%%%%%%%%%%%%%%%%%%%%%%%%%%%%%%%%%%%%%%%%%%%%%%%%%%%%%%%%%%%%%%%%%%%%%%%%%%%%%%%%%%%%%%%%%%%%%%%%%%%%%%%%%%%%%%%%%%%%%%%%%
\begin{table}[t!]
\begin{tabular}{|c||c|c|c|c|c|c|c|c|}
\hline

& $\Delta_{\textrm{sol}}^2$ & $\Delta_{\textrm{atm}}^2$  &~$\sin^2\theta_{12}$  ~&~ $\sin^2\theta_{23}$ ~&~ $\sin^2\theta_{13}$ ~&~$\delta_{CP}$ ~& ~~$\alpha_{1}$~~ & ~~$\alpha_{2}$~~ \\
&~ $[10^{-5} \,\textrm{eV}^2]$~~&~~$[10^{-3} \,\textrm{eV}^2]$ ~~&               &               &               &              &              &\\\hline\hline
Normal Hierarchy            & 7.42 & 2.514 &0.304& 0.570& 0.02221 &$195^\circ$ & 0 & 0 \\\hline
Inverted Hierarchy          & 7.42 & 2.497 &0.304& 0.575& 0.02240 &$286^\circ$ & 0 & 0 \\\hline
\end{tabular}
\caption{Best fit values of oscillation parameters for normal and inverted hierarchies used in our analysis.}
\label{table:osc-param-max-min}
\end{table}
%%%%%%%%%%%%%%%%%%%%%%%%%%%%%%%%%%%%%%%%%%%%%%%%%%%%%%%%%%%%%%%%%%%%%%%%%%%%%%%%%%%%%%%%%%%%%%%%%%%%%%%%%%%%%%%%%%%%%%%%%%%%%

%%%\noindent \underline{\it Non-unitarity}:
Constraint from non-unitarity can simply be obtained by considering the hermitian matrix $F\equiv m_D M^{-1}$.
Combining several experimental results~\cite{Fernandez-Martinez:2016lgt},
the upper bounds are given by~\cite{Agostinho:2017wfs}:
\begin{align}
|FF^\dag|\le  
\left[\begin{array}{ccc} 
2.5\times 10^{-3} & 2.4\times 10^{-5}  & 2.7\times 10^{-3}  \\
2.4\times 10^{-5}  & 4.0\times 10^{-4}  & 1.2\times 10^{-3}  \\
2.7\times 10^{-3}  & 1.2\times 10^{-3}  & 5.6\times 10^{-3} \\
 \end{array}\right].
\end{align} 
%%%
Since in our case $F$ is of the order $10^{-3}$, this constraint can be evaded at any time.
Furthermore, we expect the observed mixing is induced from $\mu_L$, therefore $m_D$ and $M$ are assumed to be diagonal for simplicity. 
%Once we conservatively take $F\approx 10^{-5}$, we find {$\mu\approx$1-10 GeV to satisfy the typical neutrino mass scale, which could be easy task.

\subsection{Lepton anomalous magnetic dipole moment and lepton flavor violations}
The muon anomalous magnetic dipole moment ($\Delta a_\mu$ or muon ($g-2$)) has been firstly reported by Brookhaven National Laboratory (BNL)~\cite{Bennett:2006fi}. They found that the muon ($g-2$) data has a discrepancy at the 3.7$\sigma$ level from the SM prediction:\
$\Delta a_\mu=(2.706\pm 0.726)\times 10^{-9}$~\cite{Blum:2018mom,Keshavarzi:2018mgv,Davier:2019can,Aoyama:2020ynm}.
Recently, new muon $g-2$ measurement in E989 experiment at Fermilab reported the new result that indicates~\cite{PhysRevLett.126.141801} 
\begin{equation}
a_\mu^{\rm FNAL} = 116592040(54) \times 10^{-11}.
\end{equation}
Combining the BNL result, anomaly of $\Delta a_\mu$ is given by
\begin{equation}
\label{eq:amu_new}
\Delta a_\mu^{\rm new} = (25.1 \pm 5.9) \times 10^{-10},
\end{equation}
where the deviation from the SM is 4.2 $\sigma$.
%

% 2009.08314
%This is because new contribution beyond the SM is highly expected.
%\footnote{There are other analyses giving slightly different estimates of the discrepancy.  For example, Ref.~\cite{Benayoun:2011mm} gives $\Delta a_\mu= (33.5\pm 8.2)\times 10^{-10}$ showing a discrepancy at the 4.1$\sigma$ level, while Ref.~\cite{fermi-lab} quotes $\Delta a_\mu= 288(63)(49)\times 10^{-11}$ indicating a 3.5$\sigma$ deviation.  In our numerical analysis, we use the result given by Ref.~\cite{Hagiwara:2011af}.}

The new contribution to the muon ($g-2$) is given by the following Yukawa Lagrangian in Eq.(\ref{yukawa}):
\begin{align}
-{\cal L}_{\Delta a_\ell}=f_{ij} \bar \ell_i P_R N_j (O_C^T)_{1,a+1}c^-_a
% (s_\gamma c^-_1 +c_\gamma c^-_2)
+ g_{ij} \bar N_i P_R \ell_j  (O_C^T)_{3,a+1}c^+_a + {\rm H.c.}
%(c_\gamma c^+_1 - s_\gamma c^+_2 )
\ ,\  {\rm a=1,2} \label{eq:mug2}
\end{align}
where $\ell$ and $N$ are assumed to be mass eigenstates. Notice here that these terms also generate LFVs.
Thus, we formulate the LFVs before discussing muon ($g-2$).

The corresponding branching ratio is given at one-loop level as follows~\cite{Lindner:2016bgg,Baek:2016kud}
\begin{align}
{\rm BR}(\ell_i\to\ell_j\gamma)
&=
\frac{48\pi^3 \alpha_{\rm em} C_{ij}}{(4\pi)^4m_{\ell_i}^2 G_F^2}
\left(
|A_{L_{ij}}|^2 + |A_{R_{ij}}|^2 
\right)
%\left(1+\frac{m_j^2}{m_i^2}\right),
\\
%%%
A_{L_{ij}}&= g^\dag_{ja} M_a f^\dag_{ai} \left[ (O_C^T)_{12}(O_C^T)_{32} I_1(M_a,m_{c_1})+  (O_C^T)_{13}(O_C^T)_{33} I_1(M_a,m_{c_2},)\right]\nn\\
&+f_{ja} f^\dag_{ai} m_{\ell_j} \left[ (O_C^T)_{12}^2 I_1(M_a,m_{c_1})+ (O_C^T)_{13}^2 I_1(M_a,m_{c_2})\right]\nn\\
&+g^\dag_{ja} g_{ai} m_{\ell_i} \left[ (O_C^T)_{32}^2 I_2(M_a,m_{c_1})+ (O_C^T)_{33}^2 I_2(M_a,m_{c_2})\right],\\
%%%
A_{R_{ij}}&= f_{ja} M_a g_{ai} \left[ (O_C^T)_{12}(O_C^T)_{32} I_1(M_a,m_{c_1})+  (O_C^T)_{13}(O_C^T)_{33} I_1(M_a,m_{c_2},)\right]
\nn\\
&+f_{ja} f^\dag_{ai} m_{\ell_i} \left[ (O_C^T)_{12}^2 I_2(M_a,m_{c_1})+ (O_C^T)_{13}^2 I_2(M_a,m_{c_2})\right]\nn\\
&+g^\dag_{ja} g_{ai} m_{\ell_j} \left[ (O_C^T)_{32}^2 I_1(M_a,m_{c_1})+ (O_C^T)_{33}^2 I_1(M_a,m_{c_2})\right],\\
I_1(m_1,m_2)&=\int[dx]_3\frac{y}{(x^2-x)m_{\ell_i}^2+x m_1^2+(y+z) m_2^2} ,\\
I_2(m_1,m_2)&=\int[dx]_3\frac{z}{(x^2-x)m_{\ell_i}^2+x m_1^2+(y+z) m_2^2},
%\frac{m_2^6 -6 m_2^4 m_1^2 + 3 m_2^2 m_1^4 +2 m_1^6+6 m_2^2 m_1^4\ln\left[\frac{m^2_2}{m^2_1}\right]}{12(m_2^2-m_1^2)^4},
\label{eq:damu1}
\end{align}
where $i,j$ runs over $e,\mu, \tau$, the fine structure constant $\alpha_{\rm em} \simeq 1/128$, the Fermi constant $G_F \simeq 1.17\times 10^{-5}$ GeV$^{-2}$, and $(C_{21}, C_{31}, C_{32}) \simeq (1, 0.1784, 0.1736)$. 
%$II(m_1,m_2) $ is derived by assuming $m_{i,j}<<M_a, m_{\eta^-}$, and notice $II(m_1,m_2) = \frac{1}{24M_a^2}$ in the limit of $M_a=m_{\eta^-}$. 
The current experimental upper bounds at 90\% C.L. are~\cite{TheMEG:2016wtm, Adam:2013mnn}
\begin{align}
{\rm BR}(\mu\to e\gamma) < 4.2\times10^{-13} ~,~
{\rm BR}(\tau\to e\gamma) < 3.3\times10^{-8} ~,~
{\rm BR}(\tau\to \mu\gamma) < 4.4\times10^{-8} ~.
\end{align}
%%%%%%%%%%%%%%%%%%%%%%%%%%%%%%%%%%%%%%%%%%%%%%%%%%
%\subsection{Anomalous Magnetic Moment of Muon and Electric Dipole Moments}
%%%%%%%%%%%%%%%%%%%%%%%%%%%%%%%%%%%%%%%%%%%%%%%%%%
Muon $g-2$ is obtained via the same interaction with LFVs, which has the form
 \begin{align}
 \Delta a_\mu \approx - \frac{m_\ell^2}{(4\pi)^2}( A_{L_{\mu\mu}}+ A_{R_{\mu\mu}}).\label{amu1L}
 \end{align}
If we consider the contribution to the electron ($g-2$), we can obtain the formula by exchanging the muon index into electron. One recent theoretical estimation~\cite{Hanneke:2008tm} of electron ($g-2$) is below\footnote{Although there is controversial nowadays. Recently, another paper~\cite{Parker:2018vye} indicates that the 
sign of electron $\Delta a_e$ might be positive.} the experimental value,
 \begin{align}
   \Delta a_e = a_e^{\rm{EXP}}-a_e^{\rm{SM}} = -(8.8\pm 3.6)\times 10^{-13}.
   \label{eq:yeg2}
 \end{align}

%where we might not disuss the electron ($g-2$) because of the sign problem. 
%%%%%%%%%%%%%%%%%%%%%%%%%%%%%%%%%%%%%%%%%%%%%%%%%%
\subsection{Flavor Conserving Leptonic $Z$ Boson Decays}\label{subsec:Zll}
%%%%%%%%%%%%%%%%%%%%%%%%%%%%%%%%%%%%%%%%%%%%%%%%%%
Here, we consider the $Z$ boson decay into two leptons through the Yukawa terms of $f$ and $g$ at one-loop level~\cite{Chiang:2017tai,Nomura:2019btk,Kumar:2020web}.
Since we assume that the off-diagonal components of $f$ and $g$ are small enough to evade bounds on LFVs, we take flavor conserving processes into account hereafter.
The new contributions to these processes are given in terms of decay rates as follows:
\begin{align}
&\Delta \Gamma(Z\to f_i\bar f_i)_{\rm new} \approx \Gamma(Z\to f_i\bar f_i)_{\rm SM+new} -  \Gamma(Z\to f_i\bar f_i)_{\rm SM},
\end{align}
where $f=\ell,\nu$ are lepton mass eigenstates. Then, each of the new contributed decay rates are given by
\begin{align}
&\Delta \Gamma(Z\to \ell_i\bar \ell_i)_{\rm new} \nn\\
&\approx \frac{m_Z}{12(4\pi)^2}\frac{g_2^2}{c_w^2}
\left[
s_w^4{\rm Re}[(O^T_{1,a+1})^2 f_{i\alpha} f^\dag_{\alpha i} I_3(M_\alpha,m_{c_a})] 
+
\left(s^2_w-\frac12\right)^2{\rm Re}[(O^T_{3,a+1})^2 g^\dag_{i\alpha} g_{\alpha i} I_3(M_\alpha,m_{c_a})]
\right], \label{eq:ztoll}\\
%%%
&\Delta \Gamma(Z\to \nu_i\bar \nu_i)_{\rm new} \approx 
\frac{m_Z}{24(4\pi)^2}\frac{g_2^2 s_w^4}{c_w^2}
{\rm Re}\left[(U^\dag_{MNS_{a i}} U_{MNS_{ia}}) f_{ia} f^\dag_{a i} (c^2_\alpha I_3(M_a,m_{h})+s^2_\alpha I_3(M_a,m_{H})\right] 
, \label{eq:zto2nu}\\
& I_3(m_1,m_2)= \int_0^1 dx (1-x)\ln[x m_1^2+(1-x)m_2^2] -\int_0^1dx\int_0^{1-x}dy \ln[(x+y) m_2^2+(1-x-y)m_1^2],
\end{align}
where $s(c)_w\equiv\sin(\cos)\theta_W\sim0.23$ stands for the sine (cosine) of the Weinberg angle.

The current bounds on the lepton-flavor-conserving $Z$ boson decay branching ratios at 95 \% CL are given by \cite{Zyla:2020zbs}:
\begin{align}
%\begin{split}
& \Delta {\rm BR}(Z\to {\rm Invisible})\approx  \sum_{i=1-3}\Delta {\rm BR}(Z\to\nu_i\bar\nu_i)< \pm5.5\times10^{-4} ,
\label{eq:zmt-con}\\
 %%%
 & \Delta {\rm BR}(Z\to e^\pm e^\mp) < \pm4.2\times10^{-5} ~,\\
& \Delta   {\rm BR}(Z\to \mu^\pm\mu^\mp) <  \pm6.6\times10^{-5} ~,
% &  \Delta   {\rm BR}(Z\to \tau^\pm\tau^\mp) <  \pm8.3\times10^{-5} ~,\label{eq:zmt-con}\\
 %%%
%&    {\rm BR}(Z\to e^\pm\mu^\mp) < 7.5\times10^{-7} ~,\
%  {\rm BR}(Z\to e^\pm\tau^\mp) < 9.8\times10^{-6} ~, \nonumber \\ 
% & {\rm BR}(Z\to \mu^\pm\tau^\mp) < 1.2\times10^{-5} ~.\label{eq:zmt-cha}
%\end{split}
\end{align}
where $\Delta {\rm BR}(Z\to f_i\bar f_j)$ ($i= j$) is defined by
\begin{align}
\Delta {\rm BR}(Z\to f_i \bar f_j)\approx 
\frac{\Delta \Gamma(Z\to \ell_i\bar \ell_i)_{\rm new}} {\Gamma_{Z}^{\rm tot}},
\end{align}
where the total $Z$ decay width $\Gamma_{Z}^{\rm tot} = 2.4952$~GeV at best fit value~\cite{Zyla:2020zbs}.
We consider all these constraints in the numerical analysis in the next section

%========================================================================================================================

\section{Results}
%%%%%%
\begin{figure}[t!]
  \includegraphics[width=77mm]{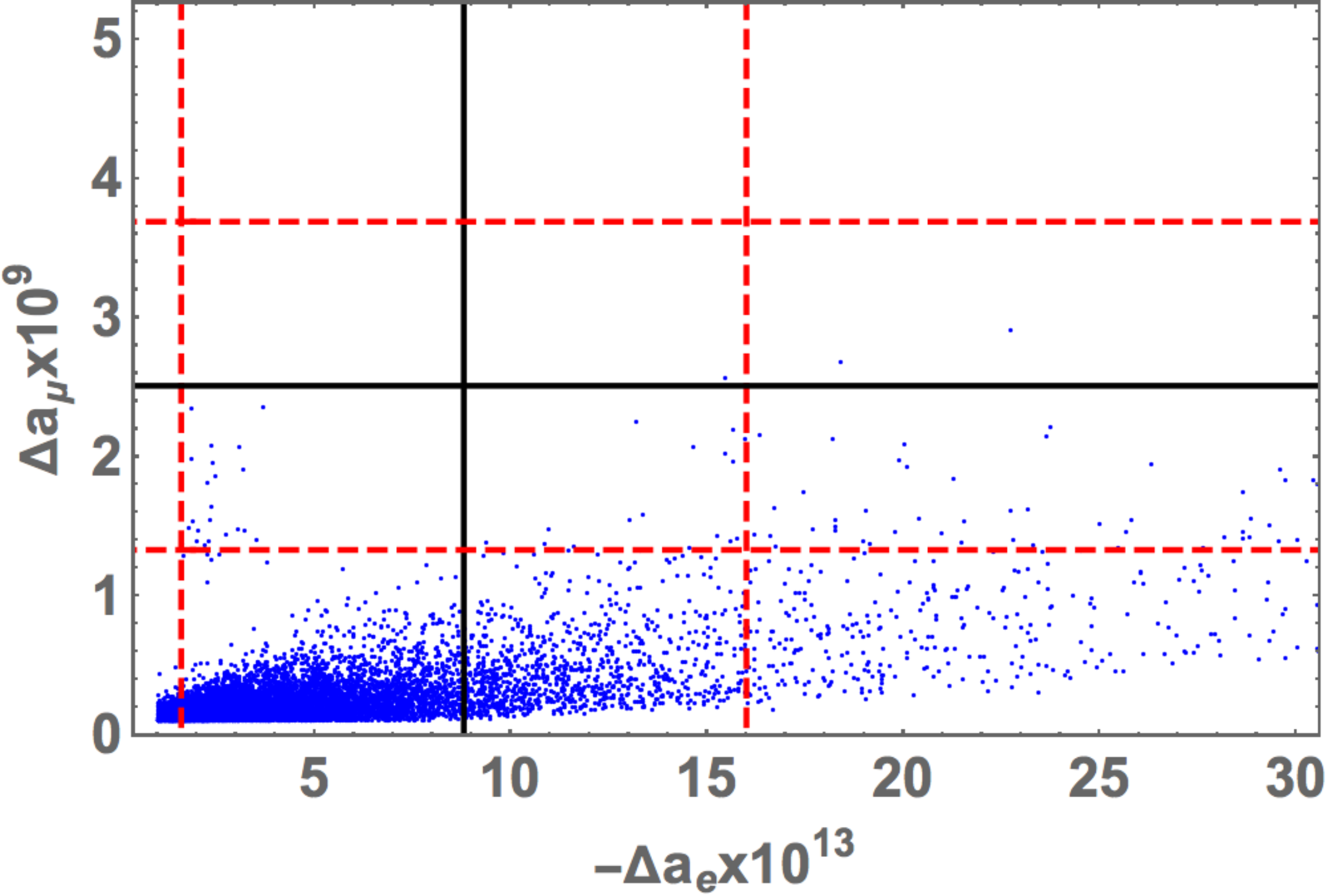}
  \includegraphics[width=77mm]{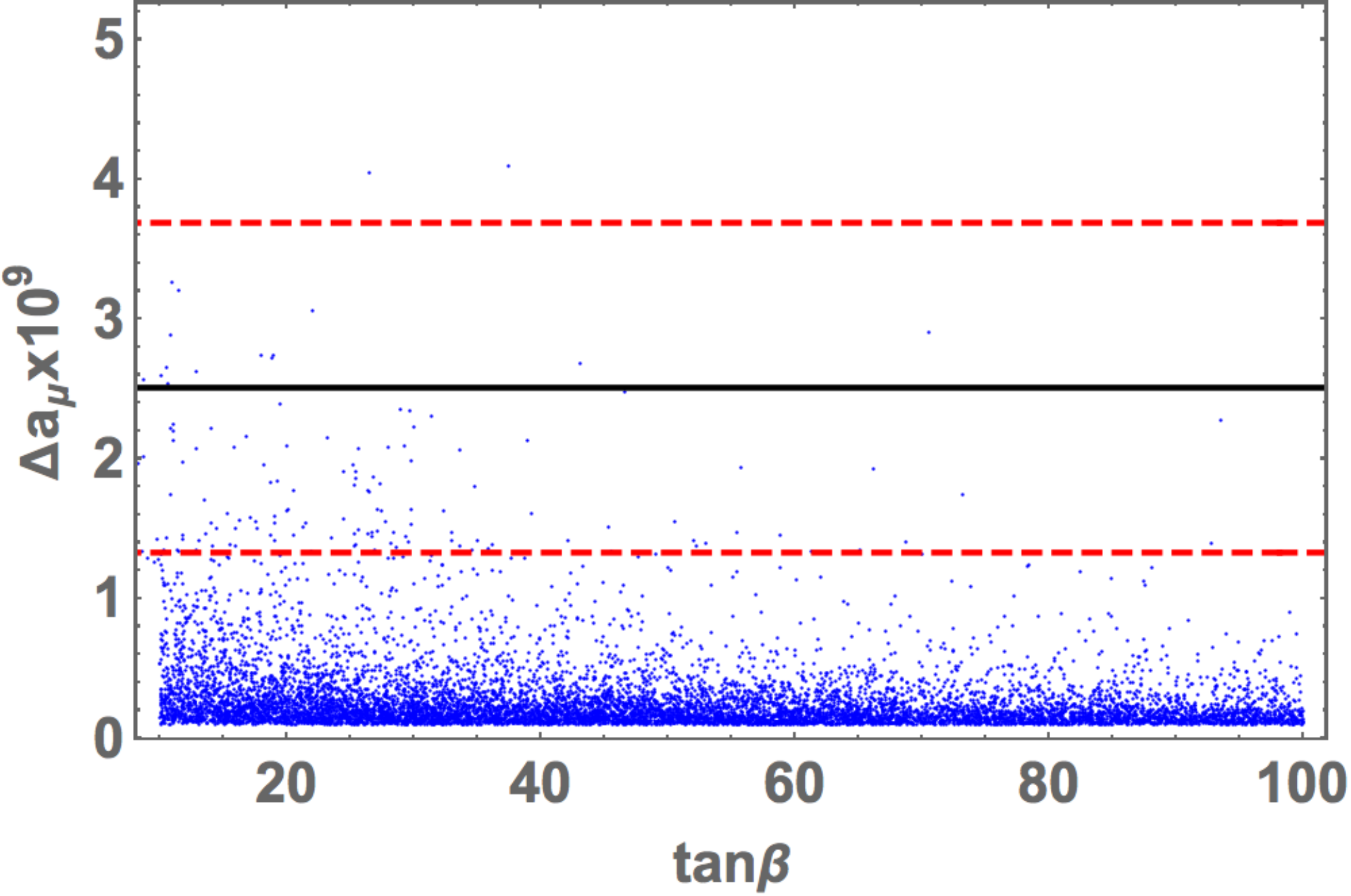}
 \\
  \includegraphics[width=77mm]{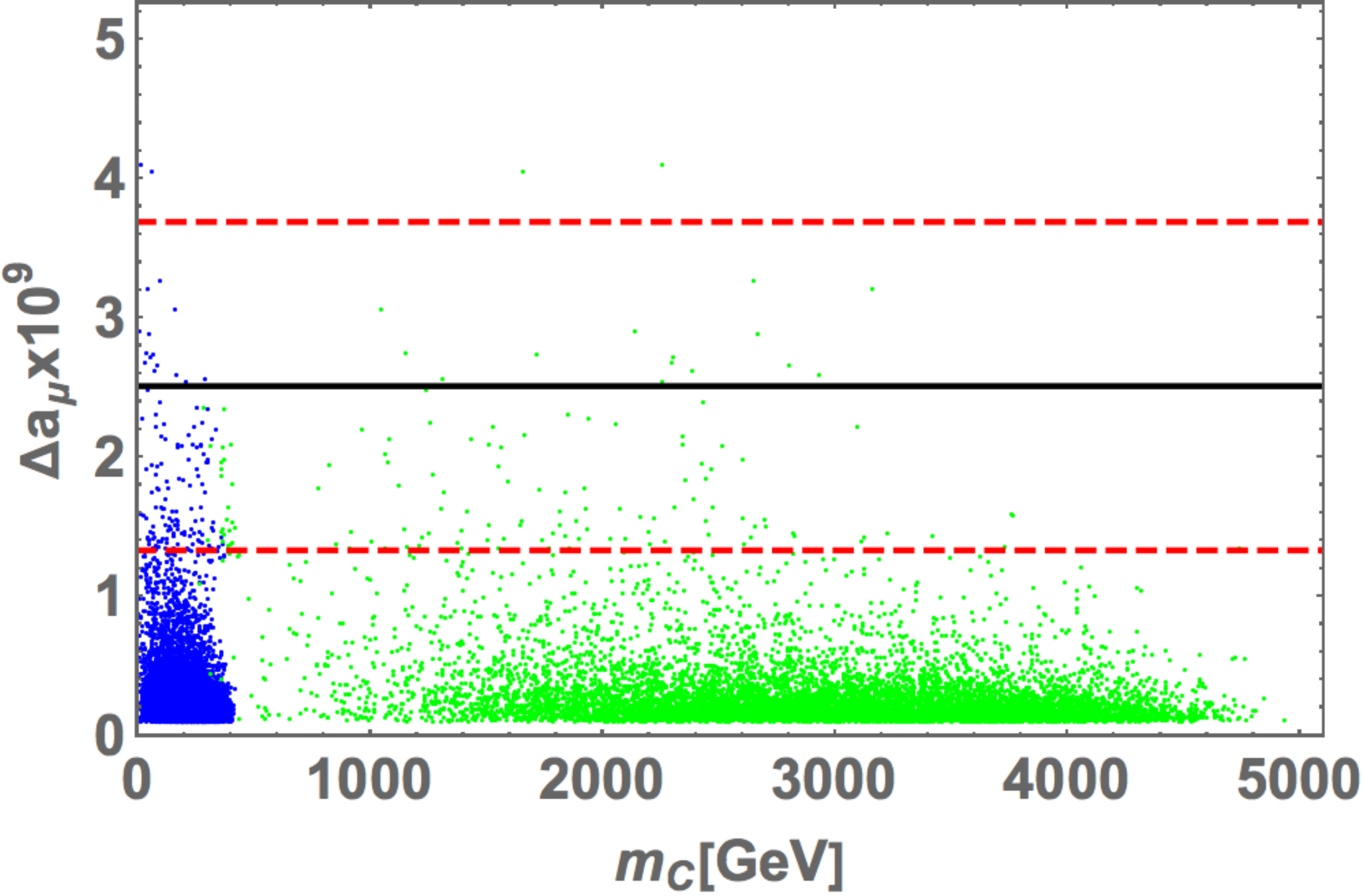}
  \includegraphics[width=77mm]{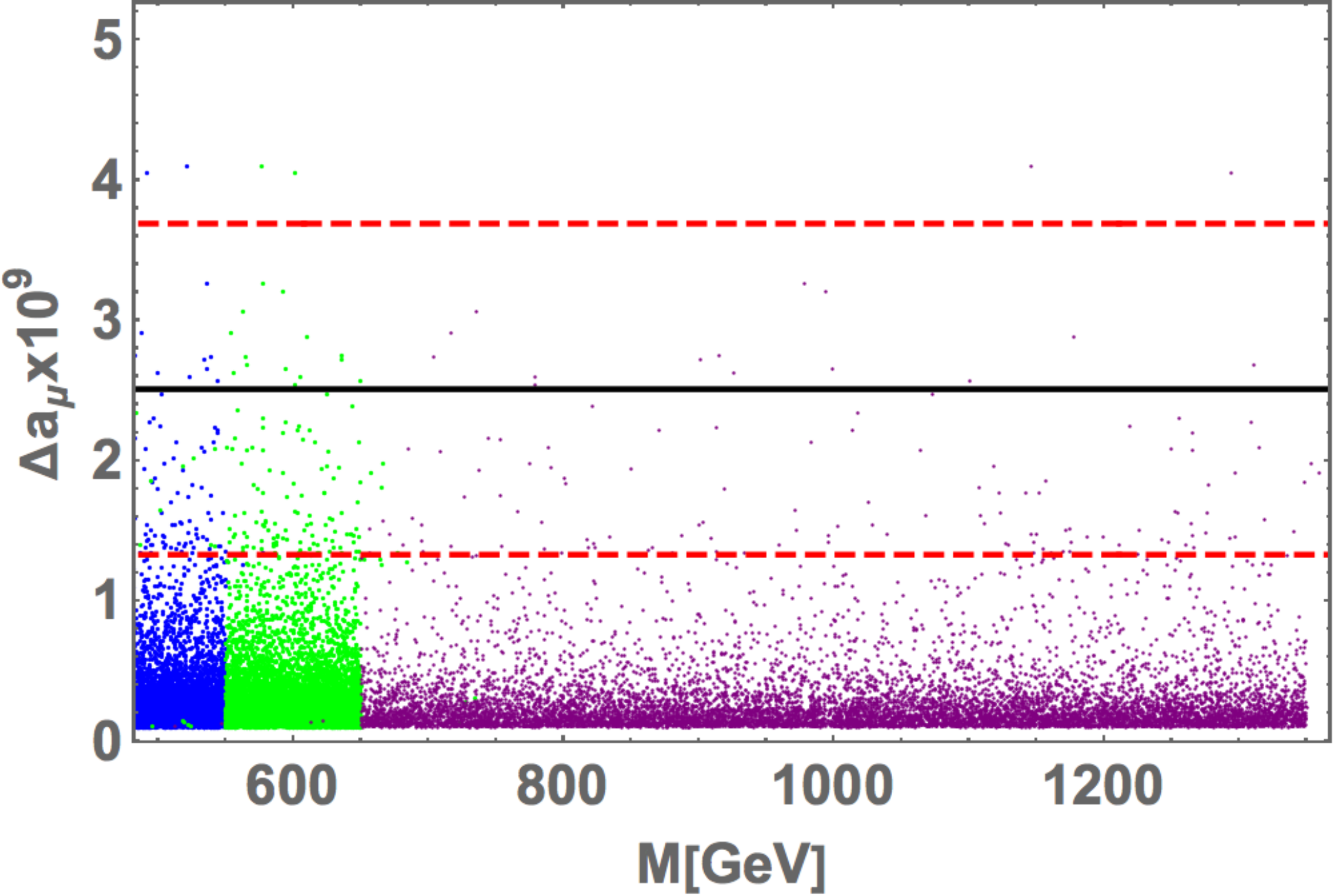}
   \caption{The scatter plots in terms of electron ($g-2$) and muon ($g-2$) for the left-top figure, muon ($g-2$) and $\tan\beta$ for the right-top one, muon ($g-2$) and charged-boson masses for left down one, and masses of $N$ for the right down one. The black lines represent the best fit values for lepton g-2, and the the red dotted line the $2\sigma$ C.L.. In the left bottom figure, the blue(green) plot is the lighter(heavier) mass of the singly-charged boson, while in the right bottom one, blue, green, and the purple color demonstrate the masses for the $N's$ in the light order.}
  \label{figs.g-2-rel}
\end{figure}
%%%%%%

\begin{figure}[t!]
  \includegraphics[width=77mm]{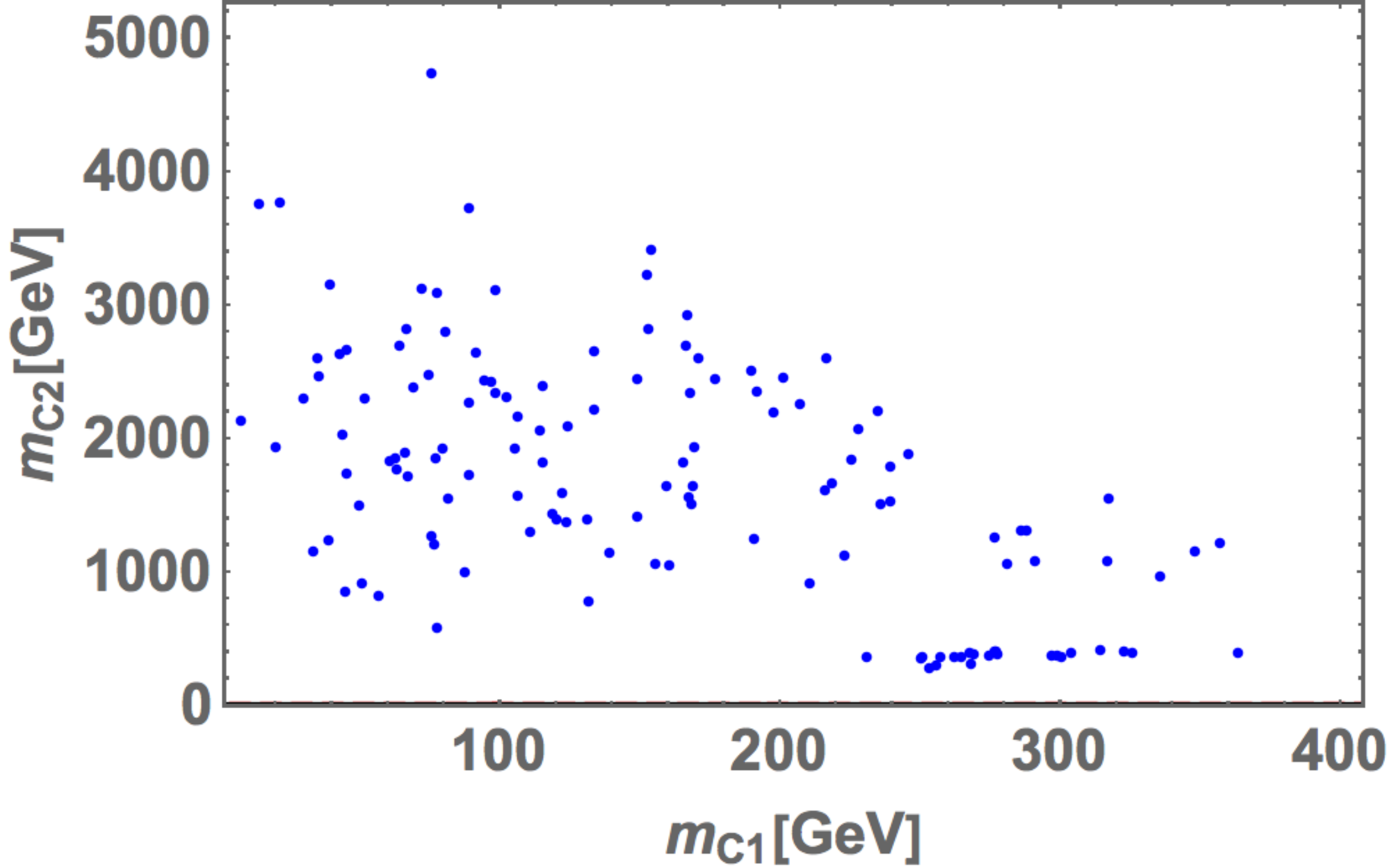}
   \caption{The scatter plots in terms of singly-charged boson masses, where these plots satisfy the muon $g-2$ at $2\sigma$ C.L..}
  \label{figs.mcgb}
\end{figure}

Here, we numerically analyze our model to search for our parameter space satisfying muon ($g-2$) or and electron g-2
as well as neutrino oscillation data.
Here, $\mu_L$ is straightforwardly rewritten in terms of our input values $m_D,\ M$ and experimental values as follows:
\begin{align}
\mu_L  =   M^*(m_D)^{-1} U_{MNS}^* D_\nu U_{MNS}^\dag (m_D^\dag)^{-1} M^\dag.
\end{align}
We scan input parameters in the following range 
\begin{align}
& \tan\beta\in[10,100], \quad v_\varphi\in [1,5]{\rm TeV},\quad [|f|_{ii},|g|_{ii}]\in[0.01,\sqrt{4\pi}],\\
&|f|_{i\neq j}=0,\quad |g|_{i\neq j} \in[10^{-5}, 10^{-2}],\quad |\lambda's| \in[10^{-3}, 1]\\
& M_1\le M_2\le M_3=[10,1500]{\rm GeV},\quad m_{c_1}\le m_{c_2}=[100,5000]{\rm GeV},
 \end{align}
 where $\lambda's$ means all the quartic couplings in Eq.~(\ref{Eq:pot}) $i,j=1,2,3$.
%%%
In Figs.~\ref{figs.g-2-rel}, we show the scatter plots in terms of electron ($g-2$) and muon ($g-2$) for the left-top figure, muon ($g-2$) and $\tan\beta$ for the right-top one, muon ($g-2$) and charged-boson masses for left down one, and masses of $N$ for the right down one. The black lines represent the best fit values for lepton $g-2$, and the the red dotted line the $2\sigma$ C.L.. In the left bottom figure, the blue(green) plot is the lighter(heavier) mass of the singly-charged boson, while in the right bottom one, blue, green, and the purple color demonstrate the masses for the $N's$ in the light order.
%%%
In Fig.~\ref{figs.mcgb}, we demonstrate scatter plots in terms of singly-charged boson masses, where these plots satisfy the muon $g-2$ at $2\sigma$ C.L..
These figures suggest that we have solutions to satisfy the muon and electron ($g-2$) at the same time at $2\sigma$ interval with the appropriate mass ranges for singly-charged bosons. The limit on the charged Higgs from the LHC collaboration depends strongly on 
$\tan\beta$~\cite{Sanyal:2019xcp,Aiko:2020ksl} and for large $\tan\beta$, only limit comes from LEP~\cite{Abbiendi:2013hk} which demands that the  charged Higgs needs to be heavier than 80 GeV.

%%%%%%
%%%%%%

%%%%%%%%%%%%%%%%%%%%%%%%%%%%%%%%%%%%%%%%%%%%%%%%%%
%\section{Collider ? \label{sec:NA}}
%%%%%%%%%%%%%%%%%%%%%%%%%%%%%%%%%%%%%%%%%%%%%%%%%%
%In this section, we present our numerical analysis

\section{Conclusions and discussions}
%%%
We have constructed a model to explain the muon and electron anomalous magnetic moments in the framework of type-X THDM with gauged $U(1)_{B-L}$ symmetry.
The neutrino mass matrix is induced at inverse seesaw mechanism that provides us milder hierarchies among neutral fermions. The sizable muon and electron ($g-2$) have been obtained at one-loop level by introducing an isospin singlet singly-charged boson that plays an role in evading the chiral suppression of these phenomenologies.
Especially, we have updated the result of muon ($g-2$) in light of the very recent report by E989 experiment at Fermilab.,
% although our situation does not change much.
%%
In our numerical analysis, we have shown our allowed regions to satisfy these lepton ($g-2$) at 2$\sigma$ interval in addition to the other constraints such as LFVs and flavor conserving leptonic $Z$ boson decays, then we have explicitly demonstrated promising benchmark points in both cases of NH and IH,
where the neutrino oscillation data can be determined by the term of $\mu_L$ that is independent of the lepton g-2. This model is verifiable in future's flavor experiments. 

%%%%%%%%%%%%%%%%%%%%%%%%%%%%%%%%%%%
\vspace{0.5cm}
\hspace{0.2cm} 

\begin{acknowledgments}
The work of T.M. is supported in part by Korean Institute Advanced Studies (KIAS) Individual Grant. No. PG073502.
The work of H.O. was supported by the Junior Research Group (JRG) Program at the Asia-Pacific Center for Theoretical
Physics (APCTP) through the Science and Technology Promotion Fund and Lottery Fund of the Korean Government and was supported by the Korean Local Governments-Gyeongsangbuk-do Province and Pohang City.
H.O. is sincerely grateful for all the KIAS members.
\end{acknowledgments}

\section*{Appendix}

Here we show a benchmark point that satisfies electron and muon ($g-2$) within 2$\sigma$ interval and also the neutrino oscillation data can be explained.
In Table~\ref{tab:benchmark} we show the input parameters. 
\begin{table}[!h]
	\centering
	\begin{tabular}{|c||c|} \hline 
		~~Input Parameter~~	  &  Value  \\ \hline 
		\rule[14pt]{0pt}{0pt}	
		$\tan\beta$&   $ 28.89$ \\ 
		\rule[14pt]{0pt}{0pt}
		$[\frac{M_1}{\rm GeV},\frac{M_2}{\rm GeV},\frac{M_3}{\rm GeV}]$ 
		&$ [377.8, 558.2, 1377]$   \\
		\rule[14pt]{0pt}{0pt}
		$[\frac{m_{c_1}}{\rm GeV},\frac{m_{c_2}}{\rm GeV}]$ 
		&$ [253.1, 283.8]$  \\
		\rule[14pt]{0pt}{0pt}
		$\frac{v_\varphi}{\rm GeV}$ 
		&2854 \\\hline
	\end{tabular}
	\caption{Numerical values of parameters and observables
		at the sample point.}
	\label{tab:benchmark}
\end{table}

The coupling matrices and charged scalar mixing matrices are,
\begin{align}
&f=
\left[\begin{array}{ccc} 
0.0316029 &0  & 0  \\
0 & 0.0527897  & 0  \\
0  &0  & -0.0232853 \\
 \end{array}\right],\quad
%%%
 g=
 \left[\begin{array}{ccc} 
-1.06494 & -0.0000385525& 0.000232422  \\
0.0000235363& 3.19761& -0.00415577 \\
-0.0000839149& -0.000650801& 0.00345861 \\
 \end{array}\right],\\
%%%
O_C&
\approx
 \left[\begin{array}{ccc} 
0.999402 & 0.0345918 & 0  \\
0.0292478 & -0.845007 & 0.533954 \\
-0.0184704 & 0.533635 & 0.845513 \\
 \end{array}\right],
 \end{align} 
These above parameters yield values of relevant observables in the allowed range and is shown in Table~\ref{tab:output}.
\begin{table}[!h]
	\centering
	\begin{tabular}{|c|c|c||c|} \hline 
		  &  Value \\ \hline 
		$\Delta a_\mu$ & $ 2.36\times 10^{-9}$\\
		\rule[14pt]{0pt}{0pt}
		$\Delta a_e$ & $-3.67\times 10^{-13}$\\
		\rule[14pt]{0pt}{0pt} 
		$[{\rm BR}(\mu\to e\gamma),{\rm BR}(\tau\to e\gamma),{\rm BR}(\tau\to \mu\gamma)]$ 
		& $[1.31\times 10^{-13},6.81\times 10^{-13}, 1.11\times 10^{-9}]$   \\
		\rule[14pt]{0pt}{0pt} 
		$[{\rm BR}(Z\to \mu\bar\mu),{\rm BR}(Z\to e\bar e),{\rm BR}(Z\to \nu\bar\nu)]$ 
		& $[2.87\times 10^{-11},4.77\times 10^{-11}, 7.64\times 10^{-11}]$   \\
		\hline
	\end{tabular}
	\caption{Obtained values of relevant observables.}
	\label{tab:output}
\end{table}

Since the neutrino oscillation data can be explained by $\mu_L$ that is independent of the lepton ($g-2$),
one can easily obtain a benchmark point for NH and IH, applying the experimental data as shown in Table~\ref{tab:benchmark}.
Then, the each $\mu_L$ is found to be
 %%%
 \begin{align}
&\frac{\mu_L}{{\rm GeV}}=10^{-9}\times
 \left[\begin{array}{ccc} 
16.6707 + 2.63336 i& -10.6146 - 6.35434 i& 172.917 + 30.8598 i \\
-10.6146 - 6.35434 i& 115.1 - 0.621073  i& -439.484 - 0.144774 i \\
172.917 + 30.8598 i& -439.484 - 0.144774  i& 2770.4 + 16.0545 i \\
 \end{array}\right]\ {\rm for\ NH},\\
 &\frac{\mu_L}{{\rm GeV}}=10^{-9}\times
 \left[\begin{array}{ccc} 
228.629  & -5.46946 - 22.3094 i & 36.7852 + 107.245 i \\
-5.46946 - 22.3094 i & 76.1496 + 1.06741 i &518.152 - 6.15507 i \\
36.7852 + 107.245 i & 518.152 - 6.15507 i & 3263.21 + 34.5101 i \\
 \end{array}\right]\  {\rm for\ IH}.
  \end{align}

% Ref Style
% Including title
%\bibliographystyle{JHEP}
%\bibliographystyle{utphys}
%\bibliography{MA4_emug2}
% \bibliographystyle{JHEP}
% \bibliographystyle{aipauth4-1}
\bibliography{MA4_emug2Notes}

\end{document}